\documentclass[11pt]{article}
\usepackage{xspace}
\usepackage{graphicx}
\usepackage{amsmath}
\usepackage{amssymb}
\usepackage{color}

\definecolor{Red}{rgb}{1,0,0}
\definecolor{Green}{rgb}{0,1,0}
\definecolor{Blue}{rgb}{0,0,1}
\definecolor{Black}{rgb}{0,0,0}

\def\beq{\begin{equation}}
\def\eeq#1{\label{#1}\end{equation}}
\def\eeqn{\end{equation}}

\def\beqa{\begin{eqnarray}}
\def\eeqa#1{\label{#1}\end{eqnarray}}
\def\eeqan{\end{eqnarray}}

\let\bar=\overbar

\def\Dslash{\not{\hbox{\kern-4pt $D$}}}
\def\dslash{\not{\hbox{\kern-2pt $\del$}}}

\def\BR{\mbox{\rm BR}}
\def\ee{e^+e^-}

\def\msb{{\bar{\ssstyle M \kern -1pt S}}}

\newcommand\vub {\ensuremath{V_{ub}}\xspace}

\def\Vub  {\ensuremath{|\vub|}\xspace}

\usepackage{relsize}
\def\babar{\mbox{\slshape B\kern-0.1em{\smaller A}\kern-0.1em
    B\kern-0.1em{\smaller A\kern-0.2em R}}}

\def\ra{\rightarrow}
\def\ellell{\ell^+\ell^-}

\def\Title#1{\begin{center} {\Large {\bf #1} } \end{center}}

\begin{document}

\Title{Indirect constraints on New Physics from the $B$-factories}

\bigskip\bigskip


\begin{raggedright}  

Alessandro Gaz\index{Gaz, A.}, {\it University of Colorado}\\

\begin{center}\emph{On the behalf of the \babar\ and Belle Collaborations.}\end{center}
\bigskip
\end{raggedright}

{\small
\begin{flushleft}
\emph{To appear in the proceedings of the Interplay between Particle and Astroparticle Physics workshop, 18 -- 22 August, 2014, held at Queen Mary University of London, UK.}
\end{flushleft}
}

\section{Introduction}

The existence of New Physics particles, with masses that can be orders of
magnitude higher than the scale of the Electroweak Symmetry breaking, can 
be probed by performing precision measurements of physics phenomena at a 
much lower energy scale.

The decays of $B$ and $D$ mesons are an excellent example of relatively 
low energy phenomena that can be sensitive to New Physics scales at the
TeV region or above, thanks to the large amount of data collected by the
\babar\ and Belle detectors at the PEP-II and KEKB accelerator facilities.

It is expected that New Physics effects will be revealed in decays that
proceed through loop or box diagrams, and thus are suppressed in the
Standard Model (SM), so that exotic particles can enter these loops and shift
the value of some of the observables from the value predicted by the 
SM. New Physics effects could also be observed at tree level in the 
hypothesis of the existence of a Higgs-like particle, whose coupling to
SM particles depends on the mass of the latter. In this case violations of
Lepton Universality could be observed.

In this contribution, I present some recent results obtained by the \babar\
and Belle Collaborations, and briefly discuss their implications for the 
indirect searches for New Physics.

\section{The \babar\ and Belle Detectors at the PEP-II and KEKB Colliders}

The \babar\ and Belle detectors, located at the PEP-II (US) and KEKB (Japan) 
$\ee$ colliders respectively, have been designed for precision studies
(particularly $CP$-violation phenomena) of the decays of $B$- and $D$-mesons, 
$\tau$ leptons, and quarkonium, and for the measurement of low-energy cross-sections
of light unflavored particles. The physics capabilities (similar for the two
detectors) include good hermeticity, high tracking efficiency and momentum
resolution, excellent vertexing resolution, high particle identification
capabilities (particularly for the $K-\pi$ separation), good energy resolution
of neutral particles in the energy range of 20 MeV to a few GeV, and high-performance
in muon reconstruction and identification.

\begin{figure}[!ht]
\begin{center}
\includegraphics[width=0.6\columnwidth]{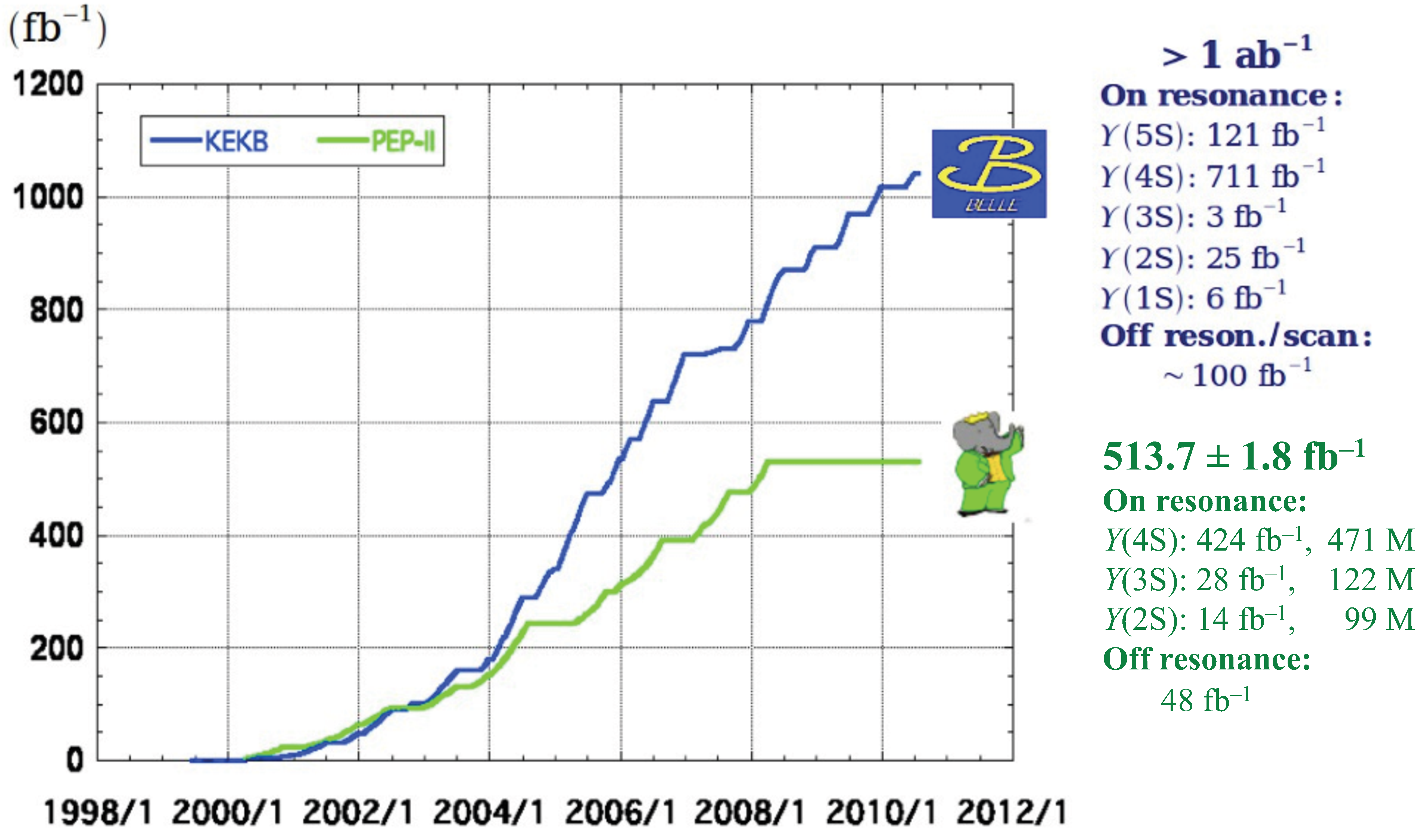}
\caption{Total integrated luminosities as a function of time for the \babar\ 
(green line) and Belle (blue) experiments. On the right-side of the
plot the breakdown of the integrated luminosity collected at different
center of mass energies.}
\label{fig:lumi}
\end{center}
\end{figure}

The data taking began in 1999 and lasted until 2008 (for \babar) and 2011 (for Belle).
Most of the data have been collected at a center of mass energy corresponding to
the mass of the $\Upsilon(4s)$ resonance, which dominantly decays to pairs of
$B^+B^-$ or $B^0\bar{B}^0$ mesons. Fig.\ref{fig:lumi} summarizes the integrated
luminosity collected by the two experiments, which represents an increase by a 
couple of orders of magnitude over the previous generation of $\ee$ colliders
at this energy.

A detailed description of the design and performance of the \babar\ and Belle 
detectors can be found in \cite{babar1,babar2} and \cite{belle}; for a summary 
of the performance of the PEP-II and KEKB accelerators, see \cite{pep2} and 
\cite{kekb}.

\section{$B$ decays proceeding through Electroweak Penguin Diagrams}

Decays of $B$-mesons that proceed through electroweak penguins and/or
box diagrams represent a very promising field in which New Physics effects
can be detected, thanks to the fact that many observables can be predicted
with high precision by the SM.
In the effective Hamiltonian:
\begin{equation}
H_{eff} = \frac{4 G_F}{\sqrt{2}}\sum_iC_i(\mu)\mathcal{O}_i
\end{equation}
short-distance effects (represented by the Wilson Coefficients $C_i(\mu)$)
can be factorized from the long-distance contributions $\mathcal{O}_i$. New
particles entering the loop could enhance/decrease the amplitude of the Wilson 
Coefficients or flip their signs, producing a significant discrepancy in the
value of one or more observables, compared to the SM expectations.

\begin{figure}[!ht]
\begin{center}
\begin{tabular}{c c}
\includegraphics[width=0.43\columnwidth]{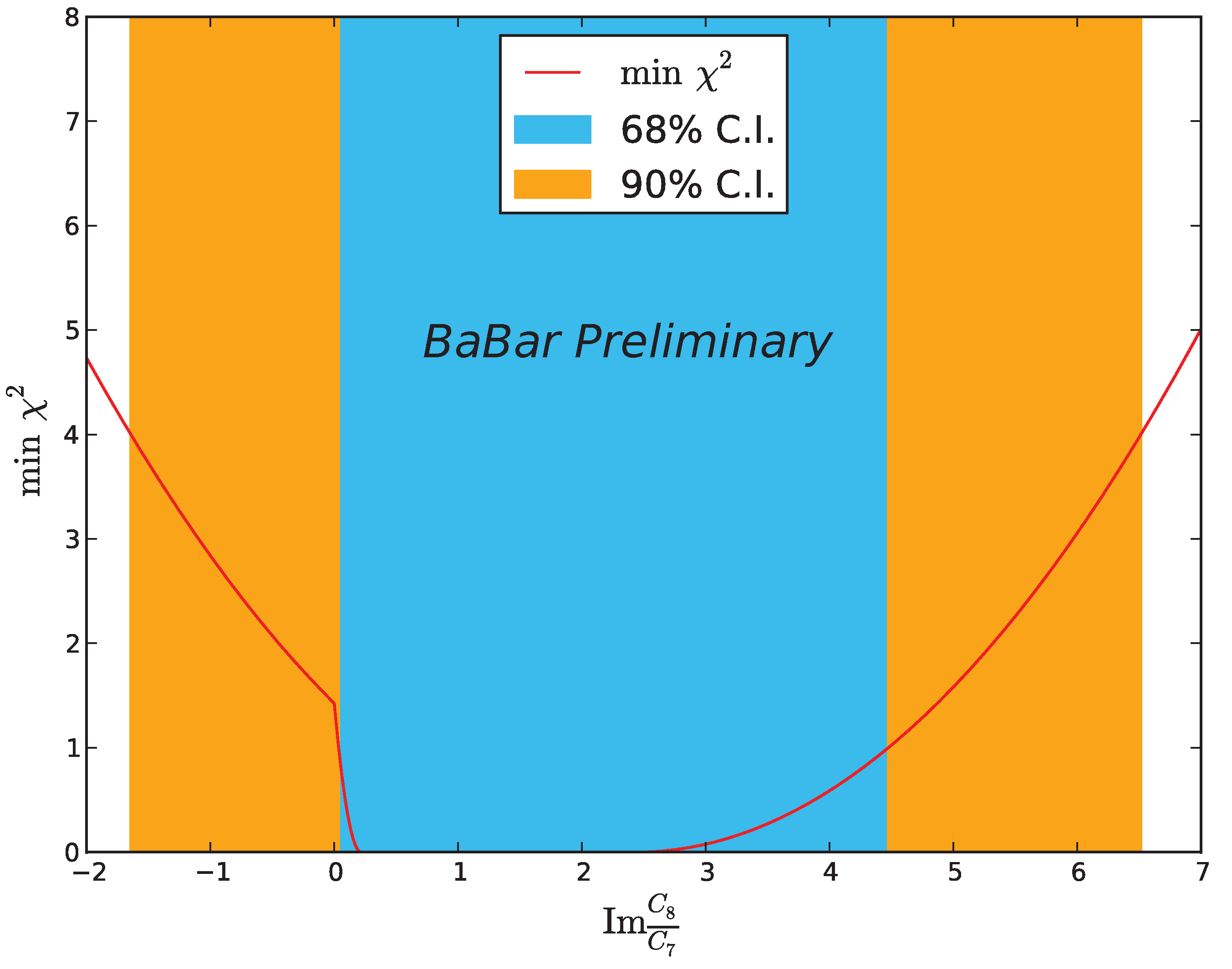} &
\includegraphics[width=0.52\columnwidth]{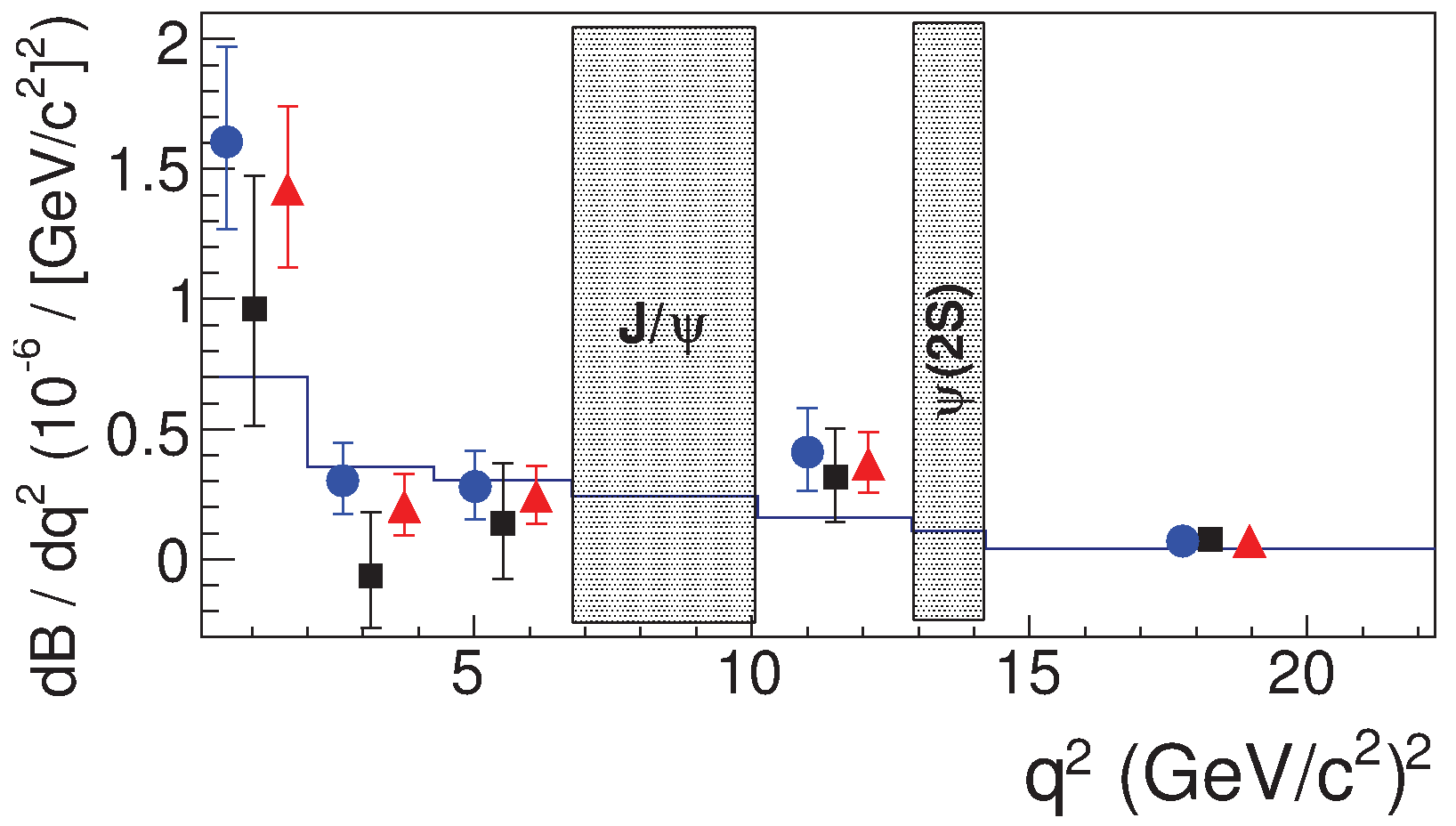} 
\end{tabular}
\caption{Left: constraints on $\Im(C_{8g}/C_{7\gamma})$ from \babar's measurement of
$\Delta A_{X_s \gamma}$ \cite{bsg_acp}. Right: differential branching fractions (as a 
function of the four-momentum transfer $q^2$) for \babar's inclusive measurement 
of $B \ra X_s \ellell$ decays \cite{bsll}. Blue dots (black squares) represent the 
$\ee$ ($\mu^+\mu^-$) results, while red triangles display the combination of the two.}
\label{fig:WilsCoeff}
\end{center}
\end{figure}

The predictions \cite{misiak} for the inclusive branching fraction of $b \ra s \gamma$ is:
\begin{equation}
\BR(B \ra X_s \gamma) = (3.15 \pm 0.23) \times 10^{-4}, \: \mbox{for } E_{\gamma} > 1.6 \mbox{ GeV},
\end{equation}
while the direct $CP$ asymmetry, defined as:
\begin{equation}
A_{CP} = \frac{\Gamma_{\bar{B}^0/B^- \ra X_s \gamma} - \Gamma_{B^0/B^+ \ra X_{\bar{s}} \gamma}}{\Gamma_{\bar{B}^0/B^- \ra X_s \gamma} + \Gamma_{B^0/B^+ \ra X_{\bar{s}} \gamma}},
\end{equation}
is expected to be compatible with 0, within a 2\% uncertainty \cite{benzke}.
\babar\ measured the inclusive $B \ra X_s \gamma$ decays using a sum of
exclusive modes \cite{bsg_acp}. The inclusive $CP$ asymmetry is measured to 
be $A_{CP} = +(1.7 \pm 1.9 \pm 1.0)\%$, consistent with the predictions.
\babar\ also measures the difference of $CP$ asymmetries for charged and
neutral modes: 
\begin{equation}
\Delta A_{X_s \gamma} = A_{B^{\pm} \ra X_s \gamma} - A_{B^0/\bar{B}^0 \ra X_s \gamma}.
\end{equation}
A non-zero $\Delta A_{X_s \gamma}$ would arise from an interference term in $A_{CP}$
that depends on the charge of the spectator quark. The measured value of
$\Delta A_{X_s \gamma} = +(5.0 \pm 3.9 \pm 1.5)\%$ is used for the first time
to constrain the imaginary part of $C_{8g}/C_{7\gamma}$, where $C_{7\gamma}$ 
($C_{8g}$) is the Wilson Coefficient corresponding to the electromagnetic 
(chromo-magnetic) dipole transition. The constraints are shown in Fig.\ref{fig:WilsCoeff}.

More constraints on potential New Physics contributions come from the measurement
of branching fractions and $CP$ asymmetry in $B \ra X_s \ellell$ decays (here and
throughout the rest of the contribution $\ell$ stands for either $e$ or $\mu$).
\babar\ measured the inclusive branching fraction and $CP$ asymmetry from a sum of
exclusive modes \cite{bsll}. The analysis is performed in different bins of
the four-momentum transfer $q^2$ (corresponding to the invariant mass of the $\ellell$
system, see Fig.\ref{fig:WilsCoeff}) and the invariant mass of the hadronic system
$X_s$. The $q^2$ regions corresponding to the invariant mass of the $J/\psi$ and $\psi(2s)$
resonances (these decays dominantly proceed through different diagrams) are excluded
from the analysis. The results are in good agreement with the SM predictions. 
As expected, also the direct $CP$ asymmetry is consistent with 0, within uncertainties:
$A_{CP} = 0.04 \pm 0.11 \pm 0.01$.

Another quantity that sparked significant interest in the past few years is the
forward-backward asymmetry $A_{FB}$ in $B \ra X_s \ellell$ decays. The angle $\theta$
that is used in the definition of the asymmetry is the angle between the $\ell^+$
($\ell^-$) and the $\bar{B}$ ($B$) meson in the $\ellell$ rest frame. New Physics
contributions could enhance/suppress or flip the sign of $A_{FB}$, compared to
what is expected in the SM. The Belle Collaboration measured $A_{FB}$ for 
the inclusive $B \ra X_s \ellell$ decays using a set of exclusive modes \cite{Afb_bsll}.
\begin{figure}[!ht]
\begin{center}
\includegraphics[width=0.45\columnwidth]{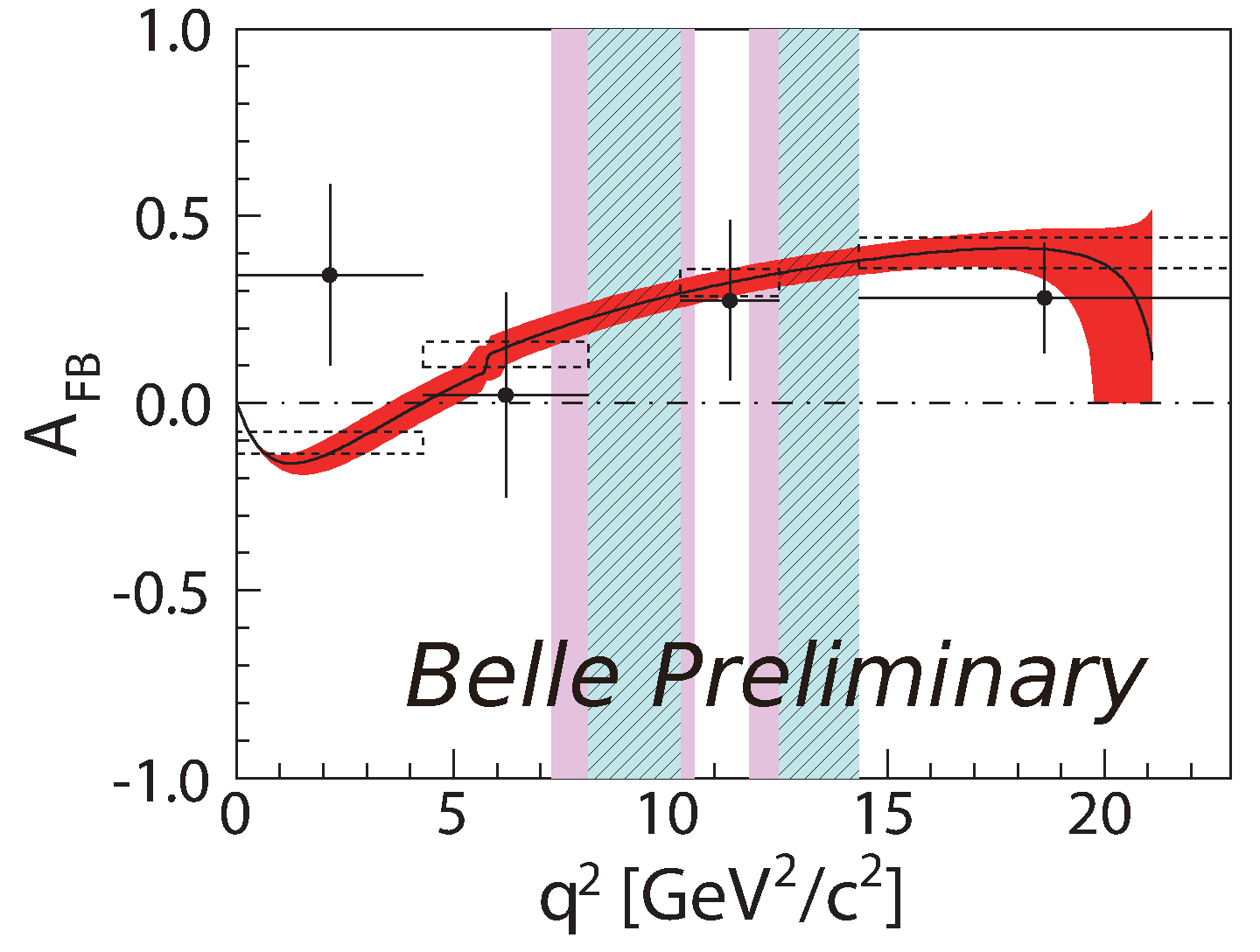} 
\caption{Forward-backward asymmetry as a function of $q^2$ for $B \ra X_s \ellell$ decays.
The red band represents the SM expectation.}
\label{fig:AFB}
\end{center}
\end{figure}
The results are reported in Fig.~\ref{fig:AFB}. The results are in good agreement
with the SM expectations, a small tension (at the 1.8$\sigma$ level) is observed
only in the low $q^2$ region, where a similar tension (not confirmed by LHCb) was 
observed in exclusive modes by \babar\ and Belle.

\section{$B$ decays to Final States containing $\tau$ leptons}

Decays with $\tau$ leptons in the final state are potentially sensitive to the
presence of Higgs-like particles, that could couple to SM particles and thus
shift branching fractions away from SM expectations and, due to the fact that
their coupling is proportional to the masses of the particles they interact
with, could lead to violation of the Lepton Universality.

In the $B^+ \ra \tau^+ \nu$ decay, an amplitude containing a charged Higgs $H^+$
could contribute along with the $W^+$ amplitude, shifting the branching fraction
away from the SM predictions. Taking the value of the CKM matrix element \Vub\ from
the CKMfitter Collaboration \cite{CKMfitter}, a branching fraction $\BR(B \ra \tau
\nu) = (0.73^{+0.12}_{-0.07}) \times 10^{-4}$ is predicted.

\begin{figure}[!ht]
\begin{center}
\begin{tabular}{c c}
\includegraphics[width=0.53\columnwidth]{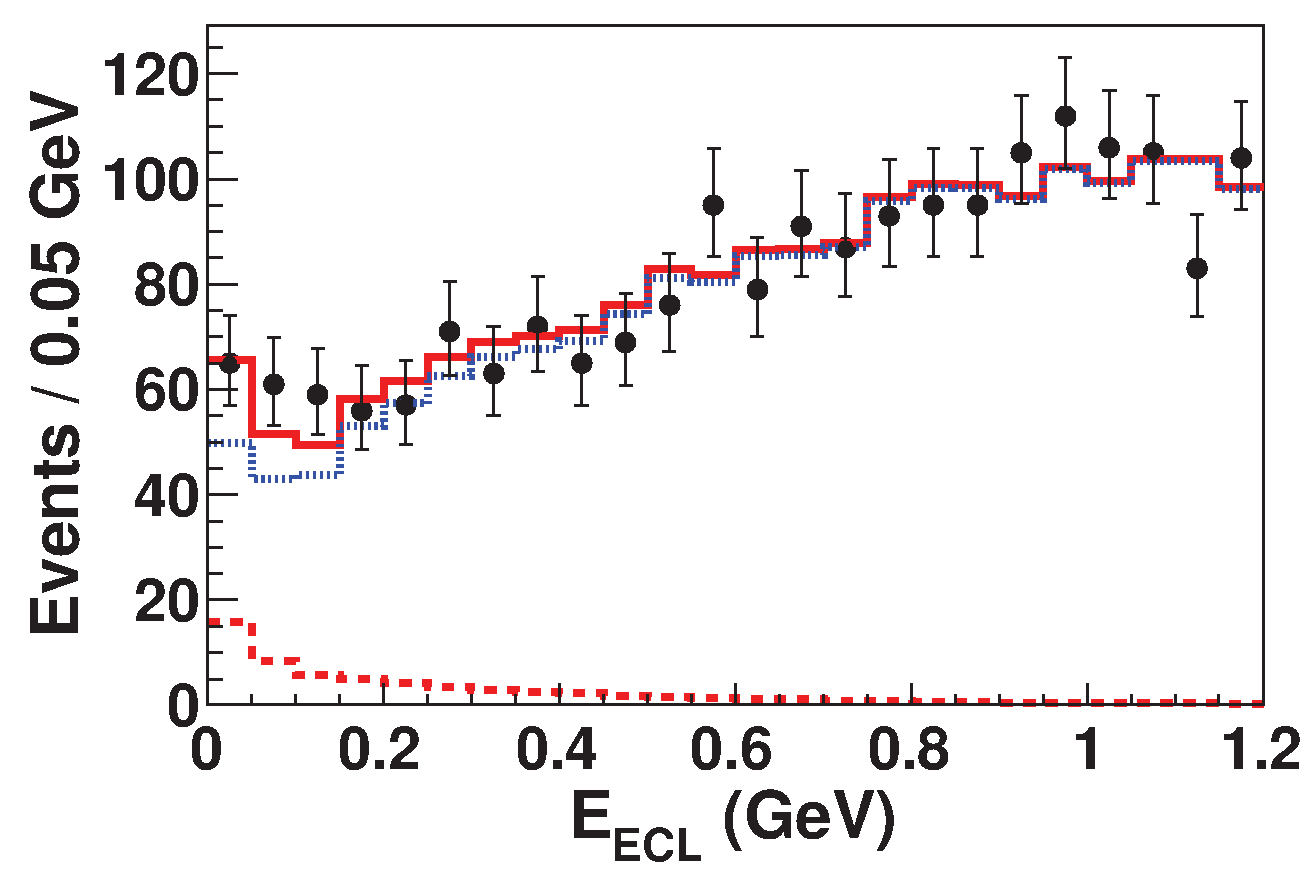} &
\includegraphics[width=0.41\columnwidth]{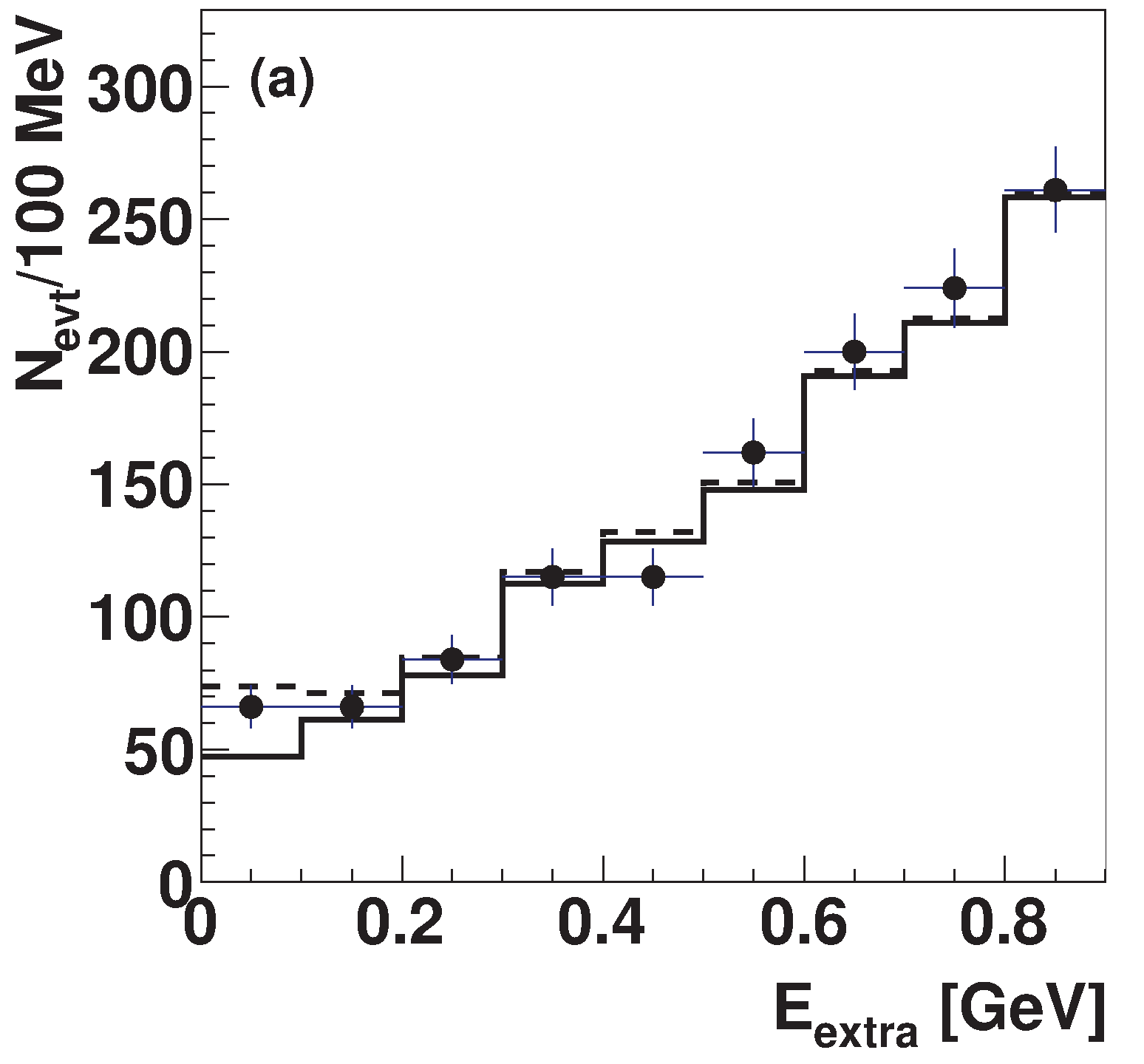} 
\end{tabular}
\caption{Distributions of the {\it extra energy} in the search for the 
$B \ra \tau \nu$ decay in Belle \cite{Btaunu_Belle} (left plot) and \babar\ 
\cite{Btaunu_BaBar} (right).}
\label{fig:Btaunu}
\end{center}
\end{figure}

The Belle Collaboration performed a measurement of this branching ratio on the
recoil of fully reconstructed $B$ decays \cite{Btaunu_Belle}. One of the two $B$ 
mesons in the event is fully reconstructed in one of many hadronic final states. 
The $B \ra \tau \nu$ decay is searched for in the rest of the event. Four $\tau$ 
decay channels have been considered: $\tau^+ \ra e^+ \nu_e \nu_{\tau}$,  
$\tau^+ \ra \mu^+ \nu_{\mu} \nu_{\tau}$, $\tau^+ \ra \pi^+ \nu_{\tau}$, and 
$\tau^+ \ra \pi^+ \pi^0 \nu_{\tau}$. In the case of a genuine signal event, no 
detectable particles should be present, besides the decay products of the fully
reconstructed $B$ candidate and the visible particles of the $\tau$ decays, so
the discriminating variable is the {\it extra energy} $E_{ECL}$ in the event.
Fig.~\ref{fig:Btaunu} shows the $E_{ECL}$ distribution for Belle's analysis:
a small excess (3.0$\sigma$ significance) is seen at very low values of $E_{ECL}$,
compatible with a $B \ra \tau \nu$ signal. The corresponding branching fraction
is: $\BR(B \ra \tau \nu) = (0.72^{+0.27}_{-0.25} \pm 0.11) \times 10^{-4}$, in very
good agreement with the SM expectations.

The \babar\ Collaboration used a very similar technique to search for the 
$B \ra \tau \nu$ decay on the recoil of fully reconstructed $B$ decays
\cite{Btaunu_BaBar}. The same final states are considered for the $\tau$ decay,
a signal excess of 3.8$\sigma$ significance is seen, which translates into a
branching fraction: $\BR(B \ra \tau \nu) = (1.83^{+0.53}_{-0.49} \pm 0.24) \times 10^{-4}$.
The tension between this result (which is in good agreement with previous
\babar\ and Belle measurements performed on the recoil of semileptonic $B$
decays) and the SM expectations is at the $2.0 \sigma$ level. Further measurements
at Belle-II will be needed in order to resolve this tension, and also the tension
between inclusive and exclusive determinations of $\Vub$.

Another class of decays potentially sensitive to the presence of charged Higgs
particles is $B \ra D^{(*)} \tau \nu$ decays, where the fact that the $H^+$ would
couple preferentially to $\tau$ leptons over the lighter $e$'s and $\mu$'s 
would lead to violations of the Lepton Universality. In order to cancel most
of the systematic uncertainties, the ratios:
\begin{equation}
R(D^{(*)}) \equiv \frac{\Gamma(B \ra D^{(*)} \tau \nu)}{\Gamma(B \ra D^{(*)} \ell \nu)}
\end{equation}
are defined. For these quantities, the SM predicts $R(D)_{SM} = 0.297 \pm 0.017$ and
$R(D^*)_{SM} = 0.252 \pm 0.003$.

\babar\ performed an analysis of these decays, considering both charged and neutral
$D$ and $D^*$ channels, using only leptonic decays of the $\tau$ \cite{BDsttaunu_BaBar}. 
In order to have good control over the backgrounds, the analysis is performed on the 
recoil of fully reconstructed $B$ decays. The discriminating variables are the missing 
mass squared and the momentum of the lepton from the $\tau$ decay. The results are in 
agreement with previous measurements and the measured branching fractions of the 
$B \ra D^{(*)} \tau \nu$ modes are somewhat higher than the expectations. The measured
values of the $R$ parameters are $R(D)_{exp} = 0.440 \pm 0.072$ and $R(D^*)_{exp} = 
0.332 \pm 0.030$, with a tension of 2.0$\sigma$ and 2.7$\sigma$ respectively with the
predicted SM values. The combined tension of the two determination with the SM
expectation is 3.4$\sigma$.

Belle performed a similar measurement on $657 \times 10^6$ $B\bar{B}$ pairs 
\cite{BDsttaunu_Belle}, and found the following values for the $R$ parameters:
$R(D^0)   = 0.70^{+0.19+0.11}_{-0.18-0.09}$, $R(D^{*0}) = 0.47^{+0.11+0.06}_{-0.10-0.07}$,
$R(D^-)   = 0.48^{+0.22+0.06}_{-0.19-0.05}$, and $R(D^{*-}) = 0.48^{+0.14+0.06}_{-0.12-0.04}$,
in agreement with \babar's results.

\begin{figure}[!ht]
\begin{center}
\begin{tabular}{c c}
\includegraphics[width=0.41\columnwidth]{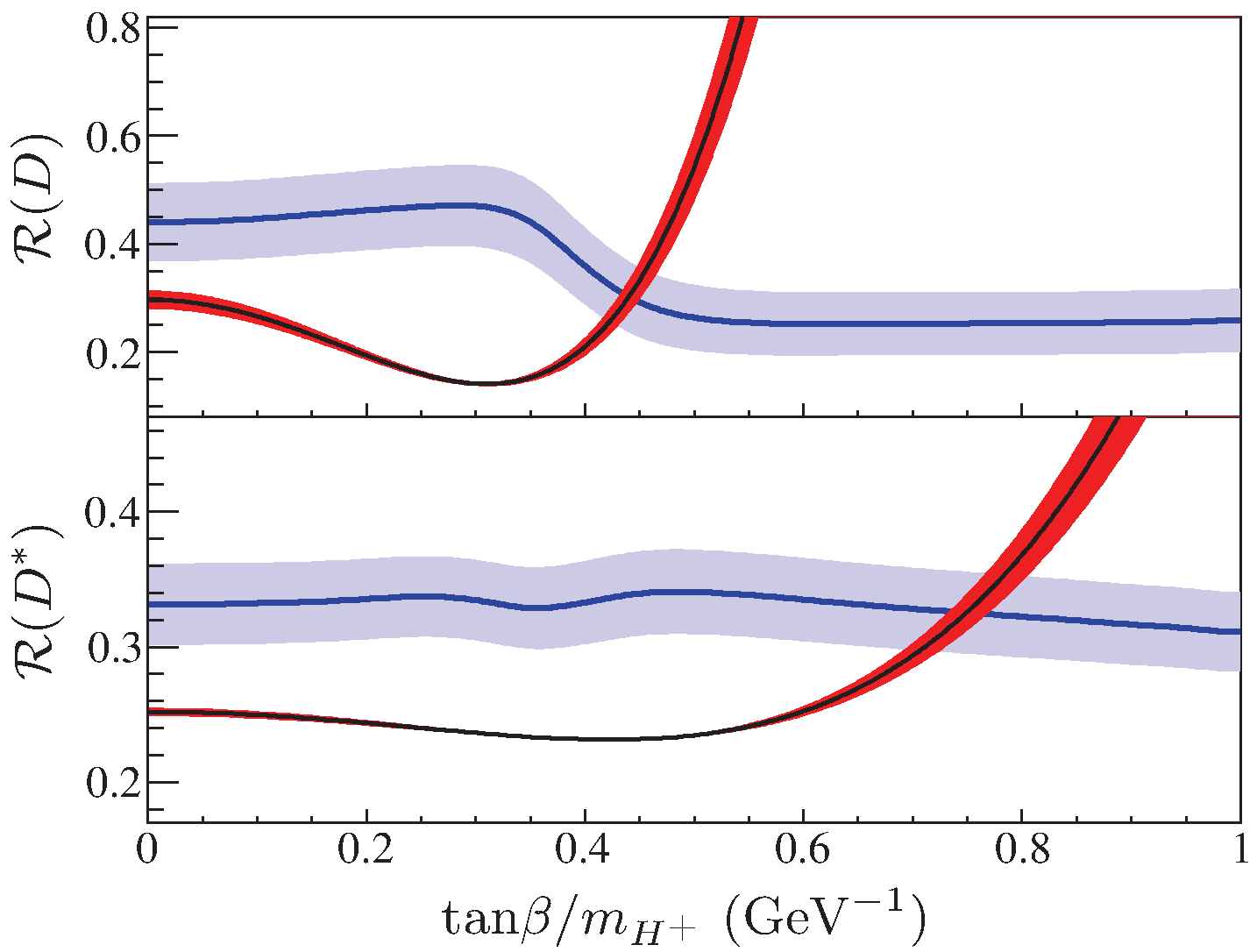} & 
\includegraphics[width=0.55\columnwidth]{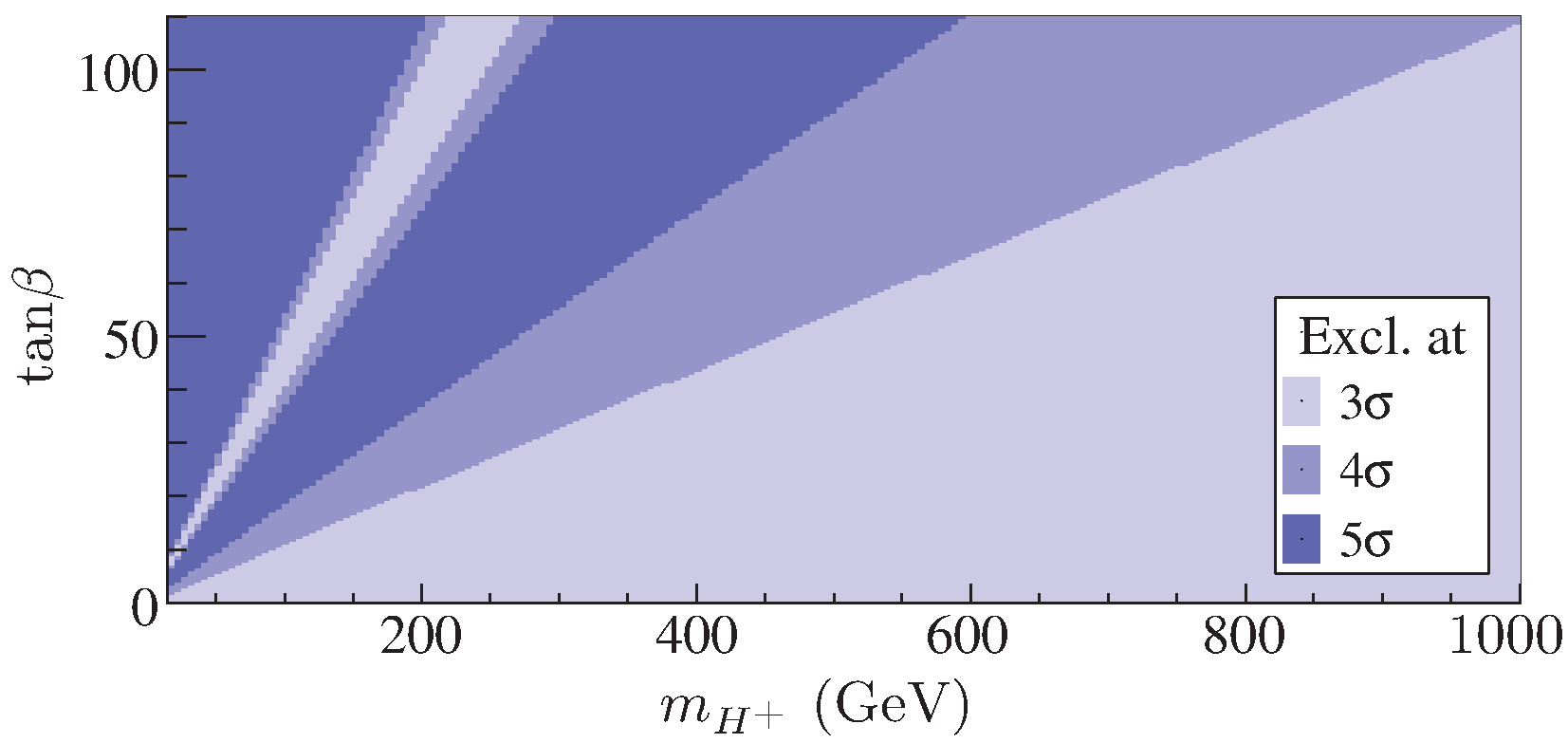} 
\end{tabular}
\caption{Left plot: \babar's experimental (blue band) and predicted (red) values 
for $R(D)$ and $R(D^*)$ as a function of the $\tan \beta / m_{H^+}$ parameter 
\cite{BDsttaunu_BaBar}. The preferred solutions for the $R(D)$ and $R(D^*)$ are 
inconsistent at the 3.0$\sigma$ level. The right plot shows the level at which 
the points in the $\tan \beta$ vs $m_{H^+}$ plane are excluded.}
\label{fig:BDsttaunu}
\end{center}
\end{figure}

The results of \babar\ are interpreted in the context of type-II 2 Higgs Doublet 
Models (2HDM), see Fig.~\ref{fig:BDsttaunu}. As it can be seen, the SM solution, 
corresponding to $(\tan \beta / m_{H^+}) = 0$, is disfavored by \babar's measurement, 
but the preferred solutions for the two independent measurements of $R(D)$ and $R(D^*)$ 
are inconsistent at the 3$\sigma$ level, so while these results are an interesting 
hint for New Physics, those also indicate that type-II 2HDM models are inadequate to 
reproduce them and more degrees of freedom would be needed if the discrepancy were
confirmed by further experiments.

\section{Charm Mixing}

The $D^0-\bar{D}^0$ mixing proceeds via the same type of box diagrams that drive 
mixing in $K^0$, $B^0_d$, and $B^0_s$ mesons. The two parameters that are commonly
used to characterize the charm mixing are $x$ and $y$:
\begin{equation}
\begin{array}{c c}
x = \frac{\Delta m}{\Gamma}, & y = \frac{\Delta \Gamma}{2 \Gamma}, 
\end{array}
\end{equation} 
where $\Delta m$ and $\Delta \Gamma$ are the mass and decay width differences of the
two mass eigenstates. Due to the importance of long-distance contributions, the SM 
cannot make reliable predictions on the values of the $x$ and $y$ parameters, besides 
the fact that they should be $\lesssim \mathcal{O}(10^{-2})$. $CP$ violation phenomena are
expected to be beyond the current experimental sensitivity, so any measurement of a
significant $CP$ asymmetry would be a clear indication of New Physics.

The Belle Collaboration measured the $D^0-\bar{D}^0$ mixing by reconstructing $D^0 \ra
K^+ \pi^-$ in their full dataset \cite{CMix_Kpi}. These decays can occur either via 
Doubly Cabibbo Suppressed (DCS) transitions, or through the Cabibbo Favored (CF) 
$D^0 \ra \bar{D}^0 \ra K^+ \pi^-$ decay that follows the oscillation. In order to 
distinguish the two processes, it is necessary to perform a time-dependent analysis 
and measure, as a function of the decay time, the ratio between the Right Sign (RS) 
and Wrong Sign (WS) events. The initial flavor of the $D^0$ is determined by the 
charge of the pion in the decay: $D^{*+} \ra D^0 \pi^+$. The results confirm the mixing
hypothesis over the no-mixing with a 5.1$\sigma$ hypothesis (first single measurement
to obtain charm mixing observation). The mixing parameters $x^{\prime}$ and $y^{\prime}$, 
where $x^{\prime} = x \cos \delta + y \sin \delta$, $y^{\prime} = y \cos \delta - x \sin \delta$, 
and $\delta$ is the strong phase between CF and DCS decays, are:
\begin{equation}
\begin{array}{c c}
  x^{\prime 2} = (0.09 \pm 0.22) \times 10^{-3}, & y^{\prime} = (4.6 \pm 3.4) \times 10^{-3}.
\end{array}
\end{equation}

\begin{figure}[!ht]
\begin{center}
\begin{tabular}{c c}
\includegraphics[width=0.44\columnwidth]{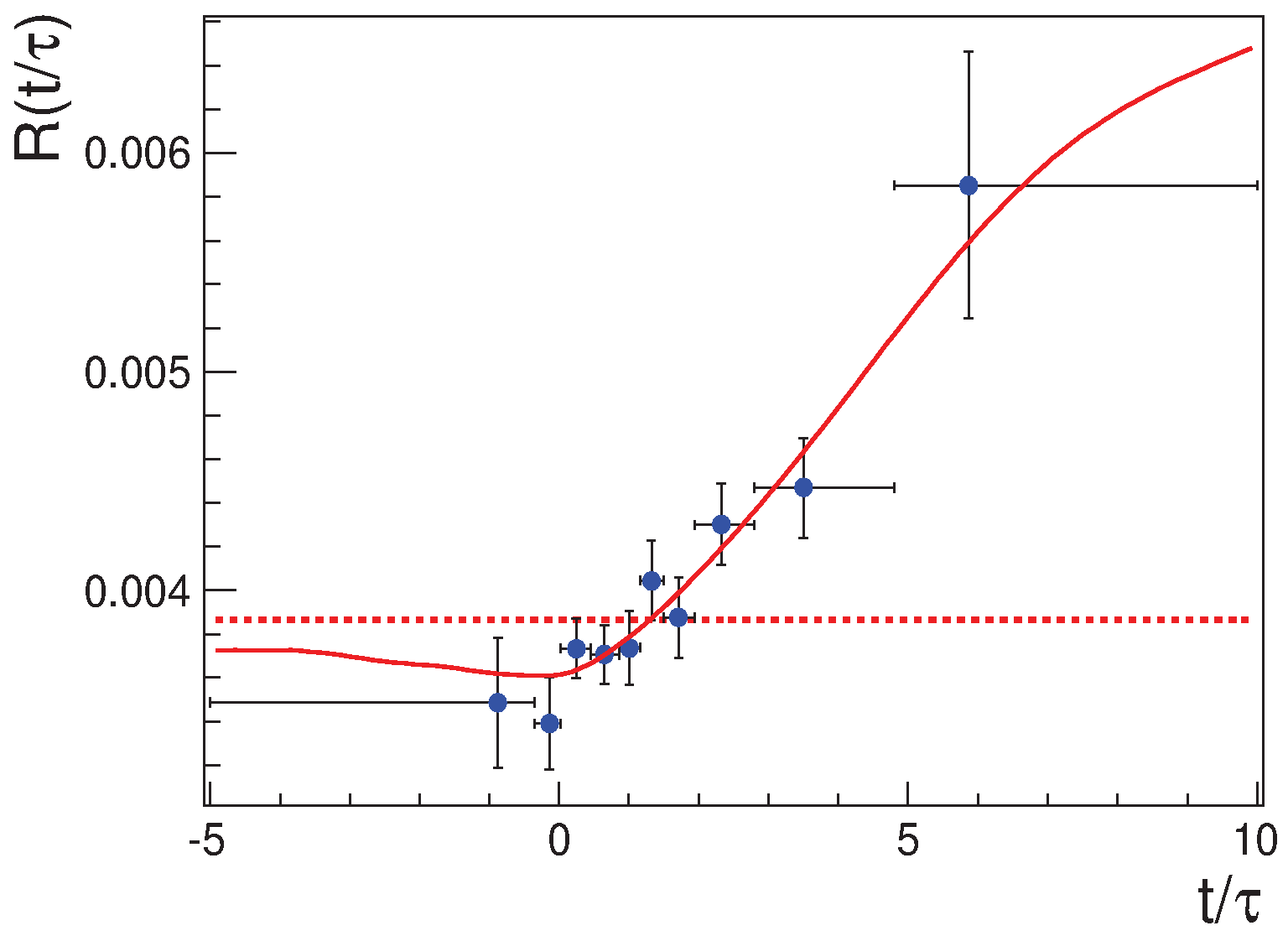} & 
\includegraphics[width=0.34\columnwidth]{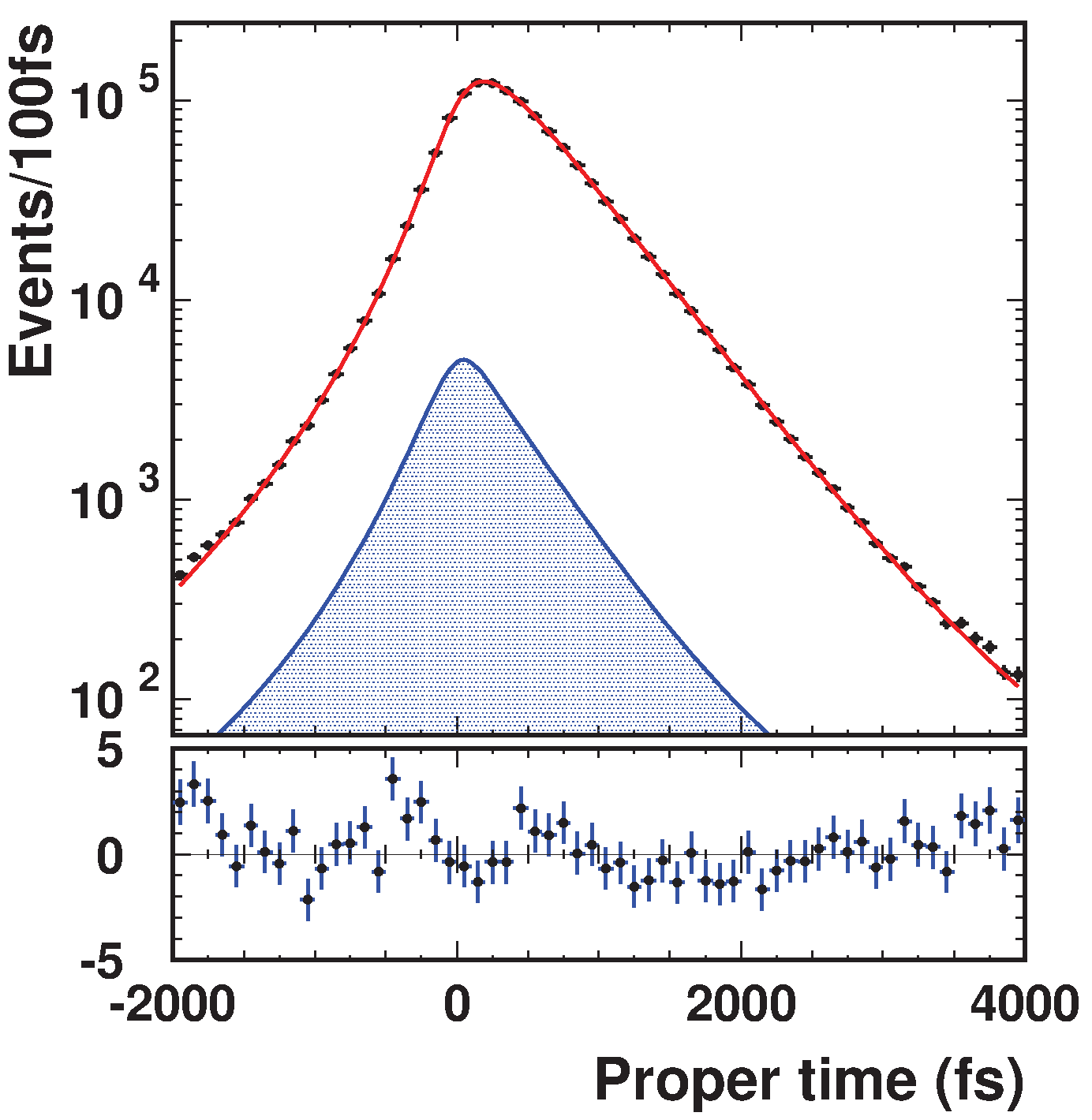} 
\end{tabular}
\caption{Proper time distribution for Belle's $K\pi$ analysis \cite{CMix_Kpi} (left) 
and $K_S\pi^+\pi^-$ Dalitz plot analysis \cite{CMix_KKpp} (right).}
\label{fig:CharmMix}
\end{center}
\end{figure}

Another method to detect charm mixing relies on comparing the $D^0$ lifetimes
in decays to $CP$-even final states and $CP$-mixed final states. The quantities
of interest are:
\begin{equation}
\begin{array}{c c}
y_{CP} = \frac{\Gamma^+ + \bar{\Gamma}^+}{2 \Gamma} - 1, & \Delta Y = \frac{\Gamma^+ - \bar{\Gamma}^+}{2 \Gamma},
\end{array}
\end{equation}
where $\Gamma^+$ ($\bar{\Gamma}^+$) is the decay width of $D^0$ ($\bar{D}^0$) to
$CP$-even final states like $K^+K^-$ or $\pi^+\pi^-$, and $\Gamma$ is the decay
width to the $CP$-mixed final states $K\pi$. \babar\ used its dataset to measure
these parameters and also to search for $CP$-violation phenomena in the interference
between mixing and decay \cite{CMix_KKpp}. The results favor the mixing hypothesis
at the 3.3$\sigma$ level and the obtained mixing parameters are:
\begin{equation}
\begin{array}{c c}
y_{CP} = (0.72 \pm 0.18 \pm 0.12)\%, & \Delta Y = (0.09 \pm 0.26 \pm 0.06)\%.
\end{array}
\end{equation}

Finally, Belle performs a time-dependent Dalitz plot analysis of $D^0 \ra K_S\pi^+\pi^-$ 
decays, exploiting the richness of the interference structure to measure the $x$ and 
$y$ parameters and to search for $CP$-violation phenomena in $D^0-\bar{D}^0$ mixing 
and in the interference between mixing and decay \cite{CMix_dalitz}. The results of the 
analysis favor the mixing hypothesis at the 2.5$\sigma$ level and no evidence of $CP$ 
violation in mixing or in the interference between mixing and decay is obtained. The fit 
results for the $x$ and $y$ parameters are:
\begin{equation}
\begin{array}{c c}
x = (0.56 \pm 0.19^{+0.03+0.06}_{-0.09-0.09})\%, & y = (0.30 \pm 0.15^{+0.04+0.03}_{-0.05-0.06})\%,
\end{array}
\end{equation}
where the first quoted error is statistical, the second is systematic, and the third is
the error associated to the amplitude model used in the analysis.

\section{Measurement of Hadronic Contributions to the anomalous
magnetic moment of the Muon}

\begin{figure}[!ht]
\begin{center}
\begin{tabular}{c c}
\includegraphics[width=0.42\columnwidth]{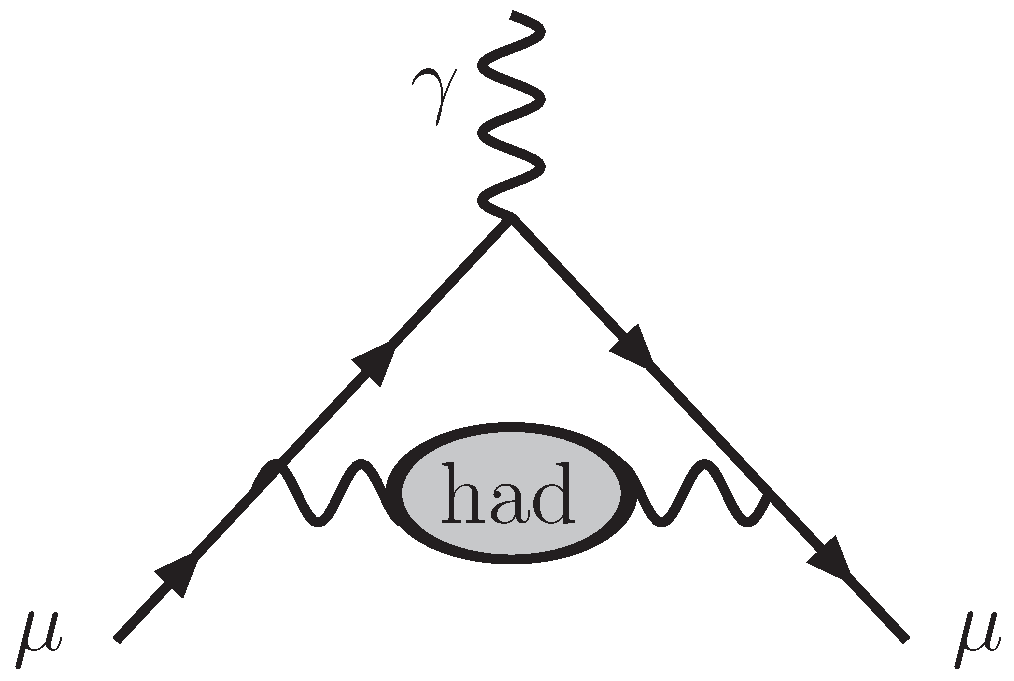} & 
\includegraphics[width=0.36\columnwidth]{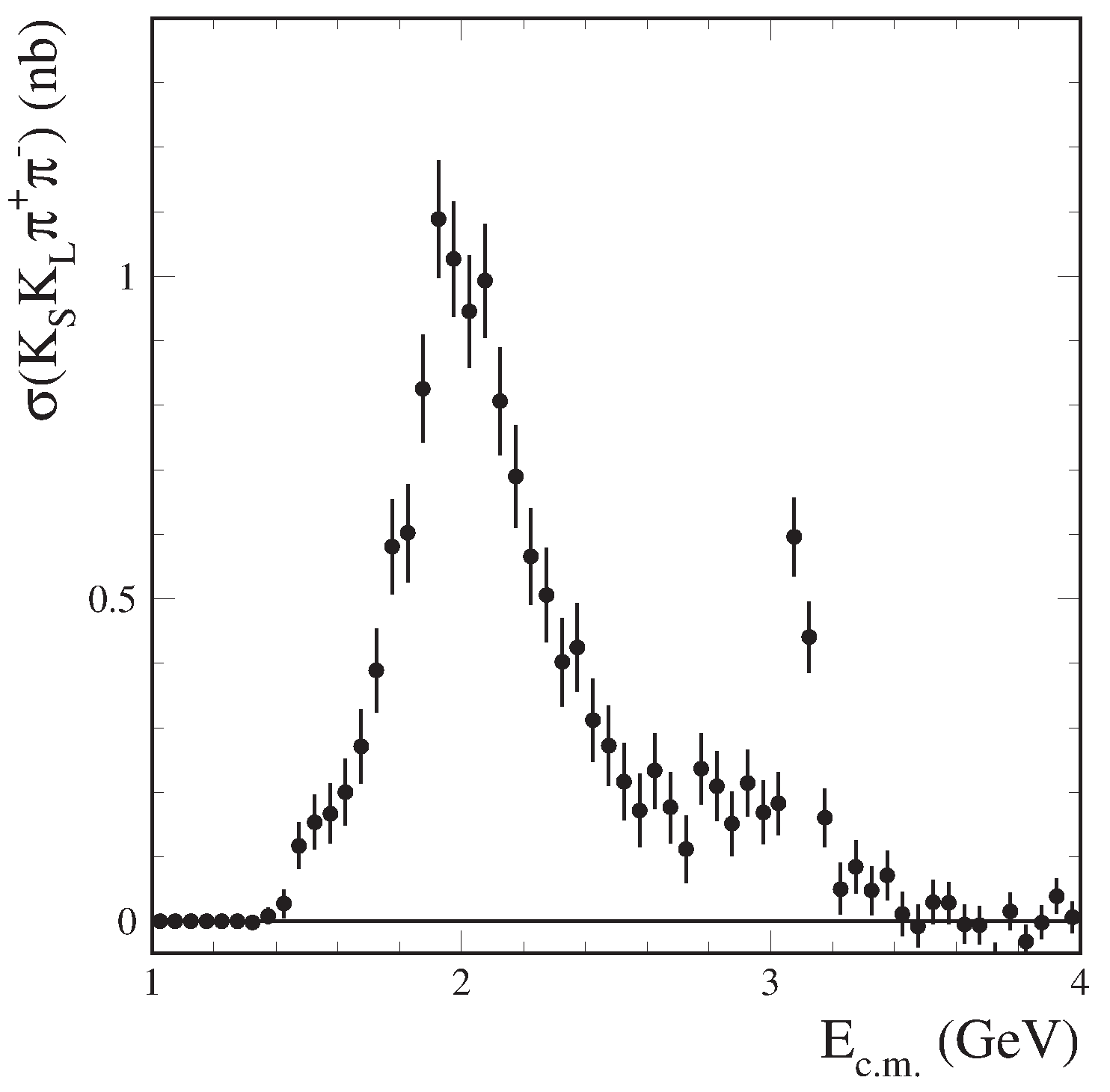} 
\end{tabular}
\caption{Left: diagram of the {\it hadronic vacuum polarization}, that contributes
to the corrections of $(g-2)_{\mu}$. Right: measurement of the $\ee \ra K_SK_L\pi^+\pi^-$
cross-section at \babar\ \cite{babar_KK}.}
\label{fig:gm2}
\end{center}
\end{figure}

There is a long standing tension between the theory predictions \cite{gm2_th}
and experimental measurements of the anomalous magnetic moment of the muon,
the so called $(g-2)_{\mu}$. While progress on the experimental side is expected
in the time scale of a few years, most of the theoretical uncertainty is dominated
by the {\it hadronic vacuum polarization} corrections, see Fig.~\ref{fig:gm2}.

The $B$-factories can play a major role in the reduction of this uncertainty, by 
measuring the cross sections of several $\ee \ra$ hadrons processes, where the invariant
mass of the hadronic systems giving sizable contributions to $(g-2)_{\mu}$ is
typically below 3 GeV. These cross-sections are connected to the hadronic vacuum
polarization through the Optical Theorem. Despite the fact that the $\ee$ collision
energy at the $B$-factories is much higher than that of the processes of interest,
the Initial State Radiation (ISR), combined with the high integrated luminosity
collected, allows to effectively perform an energy scan covering the full range
from the $\pi\pi$ threshold up to a few GeV.

The \babar\ Collaboration performed, over the last several years, a thorough
campaign of cross-section measurements in many hadronic final states. The most
important contribution to the hadronic corrections of $(g-2)_{\mu}$ comes from
the measurement of the $\pi^+\pi^-$ cross-section \cite{babar_pipi}, for
which \babar\ provides the most precise results (and a significant increase
in the discrepancy between prediction and measurement). After this measurement,
most of the uncertainty on the hadronic vacuum polarization comes from final
states with invariant mass comprised between 1.0 and 2.0 GeV. One of the
most recent \babar\ analyses on this subject \cite{babar_KK} provides the
measurement of the cross-sections of $\ee \ra K_S K_L, K_S K_L \pi^+ \pi^-, 
K_S K_S \pi^+ \pi^-, K_S K_S K^+ K^-$. The results for the $K_S K_L$ final
states are in good agreement with previous measurements, while the other
cross-sections are measured for the first time. One interesting feature of
the $K_S K_L \pi^+ \pi^-$, $K_S K_S \pi^+ \pi^-$, and $K_S K_S K^+ K^-$ 
cross-section measurement is the visible peak at a mass corresponding to 
that of the $J/\psi$ resonance, see e.g. Fig.~\ref{fig:gm2}, which
constitutes the first observation of $J/\psi$ decaying to those final
states.

\section{Summary}

A few years after the end of their data-taking the \babar\ and Belle
Collaborations continue to produce interesting results. The sensitivity of
the searches performed at the $B$-factories in many cases exceed the
direct production capabilities of the LHC, thus these searches are
complementary to those of the ATLAS and CMS Experiments. So far no significant
observation of physics beyond the Standard Model has been obtained and 
tight constraints on New Physics models have been established. There are 
nevertheless some hints of deviations that will need further investigation 
at the LHCb, Belle-II, and other Flavor Physics Experiments in the upcoming
decade.

\bigskip
\section{Acknowledgments}

The author would like to thank the IPA 2014 Organizers for the very interesting
set of topics, the right balance between talks and discussion, and the relaxed
environment in which the Conference took place.

%
%

%
%

%
%
%
%
 
\end{document}